\documentclass[10pt,twocolumn,prx,aps,amssymb,amsmath,tightenlines,nofootinbib]{revtex4} 

\usepackage{amsthm}
\usepackage{amsbsy}
\usepackage{amstext}
\usepackage{dsfont}
\usepackage{graphicx}
\usepackage[caption = false]{subfig}
\usepackage{braket}
\usepackage{tikz-cd}
\usepackage{floatrow}
\usepackage{lipsum}
\usepackage{xcolor}

\newcommand{\half}{\mbox{$\textstyle \frac{1}{2}$}}

\newcommand{\re}{\mbox{$\rm e$}}
\newcommand{\ri}{\mbox{$\rm i$}}

\newcommand{\rd}{\mbox{$\rm d$}}

\begin{document}
\title{Phase-space measurements and decoherence for angular momentum systems}

\author {\textsc{Dorje C.~Brody$^{1}$, Eva-Maria Graefe$^{2}$, and 
Rishindra Melanathuru$^{2}$}} 

\affiliation{$^{1}$School of Mathematics and Physics, University of Surrey, 
Guildford GU2 7XH, UK \\ 
$^{2}$Department of Mathematics, Imperial College London, 
London SW7 2AZ, UK 
}

\begin{abstract} 
\noindent 
The monitoring of the three independent components of the angular momentum (or spin) of a quantum system by its environment that does not isolate any preferred orientation is modelled in two different ways. One describes the dynamics by the Lindblad equation generated by three independent angular momentum operators. The other uses iterated measurements of the ``phase-space'' point on the sphere in terms of the positive operator-valued measure generated by SU(2) coherent states. In contrast to the equivalent scenario on a flat phase space, these two models give rise to subtle differences. Specifically, it is shown that the two super-operators corresponding to the two decoherence models for angular momentum systems are commutative, but their eigenvalues are different. Hence although both models give rise to phase-space decoherence, their dynamical behaviours are not equivalent. In either model, we find that the characterisation of classicality as represented by the decay rates of the elements of the density matrix (i.e. decoherence) and that as represented by the positivity of the quasiprobability distribution are not equivalent for angular momentum systems. 
\end{abstract}

\maketitle

\section{Introduction}

The loss of quantum coherence in the phase space of a quantum system can be 
modelled by assuming that the environment simultaneously monitors independent 
components of the system's phase-space variables \cite{BGM}. This differs 
fundamentally from the more familiar decoherence models arising from the 
monitoring of a single observable such as the position \cite{Zurek,JZ,Zurek22,MS}, 
which, while often investigated in phase space \cite{Davidovich1999,Zurek2001, 
Murakami2003,Ozorio2003},  need not suppress all quantum interference effects 
on phase space. In contrast, a phase-space decoherence model necessarily 
eliminates quantum interference effects on phase space entirely \cite{BGM}. 

A natural dynamical framework that captures the effect of a monitoring of 
observables in continuous time is the Lindblad equation 
\cite{Kossakowski,Lindblad,Gorini}, driven by these independent observables. 
For instance, for a quantum system defined on the real line (or in two- or 
three-dimensional space), the relevant phase-space variables are the position 
and momentum. In this case, phase-space decoherence can be modelled by a 
Lindblad equation with Lindblad operators given by the position and momentum 
observables themselves \cite{BGM,DeBievreEtAl2019,HertzDeBievre2020}. It 
has been shown in \cite{BGM} that the effect of such Lindbladian dynamics on 
the system is equivalent to that of repeated measurements of the phase-space coordinates 
described by the positive operator-valued measure (POVM) associated with 
coherent states. That is, the Lindblad equation can be unravelled 
by means of repeated phase-space POVM measurements. The outcome 
of each POVM measurement is a phase-space point, and the resulting state 
is a coherent state centred at that point. More generally, 
if the system is monitored by the environment, then we have a nonselective 
measurement, and the outcome state is the projection operator onto a coherent 
state, averaged over the phase space with respect to the Husimi distribution of 
the initial state. This process can be iterated successively to yield nontrivial 
state transformations, because no two outcome states can be orthogonal. 
The equivalence with the Lindbladian dynamics driven by position and momentum 
can be deduced from the fact that the solution to the 
Lindblad equation at the discrete times $t_k=\lambda k$, with $k=0,1,2,\ldots$, 
for a suitable constant $\lambda>0$, agrees with the outcome states of $k$ 
successive POVM measurements \cite{BGM}. This is perhaps expected, given 
that both schemes model measurements of the phase-space location of the 
system. 

In the present paper we investigate an analogous situation for quantum systems 
described in terms of angular momentum
operators, where the phase space is 
spherical. Specifically, we consider a Lindblad evolution modelling the 
simultaneous continuous measurement of the three orthogonal angular 
momentum components, in comparison to the effect of successive POVM 
measurements with respect to spin coherent states. Given the equivalence of the 
two models in the case of a flat phase space, and the intimate connection of both 
the Lindblad dynamics and the coherent states to the heat equation \cite{Hall}, 
one might expect the two phase-space decoherence models to agree. 
However, by deriving  explicit solutions for the time-dependent 
density operator in both models, 
expanded in terms of the spin-moment components, 
we reveal a discrepancy between the two descriptions. 
While both models lead to the expected decoherence effect of 
dampening the off-diagonal elements of the density matrix and driving it 
eventually to the state of complete ignorance, we find that the decay rates of 
each of the moments are different (with the exceptions of spin-$\frac{1}{2}$ 
systems). The qualitative effects of both phase-space decoherence models 
are equivalent, yet the difference 
in the decay rates should be experimentally measurable, allowing an 
experimental investigation into whether either of the two models introduced 
here might capture the effect of an environment that does not isolate any 
preferred spin orientation. 

The paper is organised as follows. In Section~\ref{sec:2} we investigate the 
continuous-time phase-space decoherence model given by the Lindblad 
equation for the density matrix, and present the solution. In Section~\ref{sec:3} 
we compare this to the effect of phase-space POVM measurements on the 
density matrix. In Section~\ref{sec:4} we compare the corresponding 
decoherence rates for the two decoherence models, and show that in general 
they do not agree. In Section~\ref{sec:5} we recast these results in terms of 
phase-space quasidistribution functions, working out explicit representations 
of the solution to the dynamical equation. We conclude in Section~\ref{sec:6} 
with a brief summary and discussion.  

\section{Continuous-time spin monitoring}
\label{sec:2}
 
For the purpose of modelling the continuous-time monitoring of the three 
independent spin components by the environment, let us consider an 
angular momentum system whose state obeys the Lindblad equation 
\begin{eqnarray}
\frac{\rd\hat{\rho}}{\rd t} &=& \gamma \left( 
\hat{J}_{x} \hat{\rho} \hat{J}_{x} + \hat{J}_{y} \hat{\rho} \hat{J}_{y} + 
\hat{J}_{z} \hat{\rho} \hat{J}_{z}  - \frac{1}{2} \left( \hat{J}_{x}^{2} \hat{\rho} 
+ \hat{\rho} \hat{J}_{x}^{2} \right) \right. \nonumber \\ && \left. 
- \frac{1}{2} \left( \hat{J}_{y}^{2} \hat{\rho} 
+ \hat{\rho} \hat{J}_{y}^{2} \right) - \frac{1}{2} \left( \hat{J}_{z}^{2} \hat{\rho} 
+ \hat{\rho} \hat{J}_{z}^{2} \right) \right), 
\label{eq:1} 
\end{eqnarray}  
where $\gamma>0$ is a parameter that determines the decoherence rate, and 
the three spin operators ${\hat J}_x$, ${\hat J}_y$, and ${\hat J}_z$ 
driving the dynamics satisfy the usual $\mathsf{SU}(2)$ commutation relations 
$[{\hat J}_x,{\hat J}_y]=\ri{\hat J}_z$, $[{\hat J}_y,{\hat J}_z]=\ri{\hat J}_x$, 
and $[{\hat J}_z,{\hat J}_x]=\ri{\hat J}_y$. 
This Lindblad equation has been investigated in the literature in the context 
of depolarisation dynamics \cite{RL}. It can be derived, for example, using 
a microscopic model of a spin interacting with a fluctuating magnetic field
\cite{DM}, but also based on other setups \cite{TS,KRS,AJD}. (In Appendix 
\ref{app_LB} we provide another derivation of a microscopic model 
of a particle in an environment that monitors its angular momentum components.) 
An alternative way of expressing the dynamical equation is to use the spin 
raising and lowering operators ${\hat J}_\pm={\hat J}_x\pm\ri {\hat J}_y$. 
In terms of these operators, 
the dynamical equation (\ref{eq:1}) can be expressed in the form 
\begin{eqnarray}
\frac{\rd\hat{\rho}}{\rd t} = \gamma \left( \tfrac{1}{2} 
\hat{J}_{+} \hat{\rho} \hat{J}_{-} + \tfrac{1}{2} \hat{J}_{-} \hat{\rho} \hat{J}_{+}
+ \hat{J}_{z} \hat{\rho} \hat{J}_{z} - J(J+1) \hat{\rho} \right) .
\label{eq:2} 
\end{eqnarray}

Although the solution to the dynamical equation (\ref{eq:1}) is known in the 
literature (see \cite{DM} and references cited therein), we shall derive it 
here so that comparison with the POVM measurement becomes more 
transparent. To solve the Lindblad equation we find it convenient to use 
irreducible tensor operators $\{\hat T^{\,J}_{L,k}\}$, which constitute a 
suitable basis for analysing spin systems 
\cite{Varshalovich,Racah1942,Edmonds}. These operators satisfy the 
commutation relations 
\begin{eqnarray}
[{\hat J}_z,{\hat T}^{\,J}_{L,k}] = k\, {\hat T}^{\,J}_{L,k} 
\end{eqnarray}
and
\begin{eqnarray}
[{\hat J}_\pm,{\hat T}^{\,J}_{L,k}] = 
\sqrt{(L \mp k)(L \pm k + 1)}\, {\hat T}^{\,J}_{L, k\pm 1} \, ,
\end{eqnarray}
and form a 
Hilbert-Schmidt orthonormal operator basis in the sense that 
\begin{eqnarray}
{\rm tr} \left({\hat T}^{\,J}_{L,k} ({\hat T}^{\,J}_{L',k'})^\dagger\right) 
= \delta_{L L'} \, \delta_{k k'},
\end{eqnarray}
where $({\hat T}^{\,J}_{L,k})^{\dagger} = (-1)^k\,\hat T^{\,J}_{L,-k}$. 
The index $L$ runs from $0$ to $2J$, while $k$ runs from $-L$ to 
$L$. The index $J$ on the other hand is fixed, and acts as a placeholder to 
indicate the total spin of the system. 
As an example, for a spin-1 system, we have the nine basis elements  
\begin{align}
\hat{T}^{\,1}_{0,0} &= \frac{1}{\sqrt{3}}\mathds{1},\\
\hat{T}^{\,1}_{1,0} &= \frac{1}{\sqrt{2}}\hat J_z,\\
\hat{T}^{\,1}_{1,\pm1} &= \mp \frac{1}{2}\hat J_\pm,\\
\hat{T}^{\,1}_{2,0} &= \frac{1}{\sqrt{6}}(3\hat J_z^2-\hat J^2),\\
\hat{T}^{\,1}_{2,\pm1} &= \mp \frac{1}{2}(\hat J_\pm\hat J_z+\hat J_z\hat J_\pm),\\
\hat{T}^{\,1}_{2,\pm2} &= \frac{1}{2}\hat J_\pm^2 .
\end{align}
These can be used to expand any $3 \times 3$ observable. More generally,
the matrix elements of the irreducible 
tensors in the standard $\hat J_z$-bases are given by
\begin{eqnarray}
\langle J,m'|\hat T^{J}_{L,k}|J,m\rangle=\sqrt{\frac{2L+1}{2J+1}}\, 
C_{JmLk}^{Jm'} \, ,
\end{eqnarray}
where the $C_{j_1m_1j_2m_2}^{JM}$ denote the Clebsch-Gordan coefficients. 

In terms of the irreducible tensor operators the initial state of the system can be 
expanded according to 
\begin{eqnarray}
\hat\rho = \sum_{L=0}^{2J}\sum_{k=-L}^{L} \rho_{Lk}\, 
{\hat T}^{\,J}_{L,k}, 
\quad
\rho_{L,k} = {\rm tr}\!\left(({\hat T}^{\,J}_{L,k})^{\dagger}\,{\hat\rho}\right).
\label{eq:3} 
\end{eqnarray}
The expansion coefficients are often referred to as the monopole moment for 
$\rho_{00}$, the vector moments or orientations for $\rho_{1k}$, $k=-1, 0, 1$, 
the quadrupole moment or alignment for $\rho_{2k}$, $k=-2,\ldots,2$, and so on 
\cite{VMK,Blum}. 

What makes the irreducible tensor operator basis particularly well suited for our 
purposes is that the $\{\hat T^{\,J}_{L,k}\}$ are
eigenoperators of the Lindblad equation 
(\ref{eq:1}). Specifically, writing the right side of (\ref{eq:1}) in the form  
\begin{eqnarray}
\mathcal{L}({\hat\rho})=-\frac{1}{2}\gamma\sum_{i=x,y,z}[\hat{J}_i,[\hat{J}_i,\,
{\hat\rho}]],
\end{eqnarray}
we have (see \cite{Racah1942,Edmonds}) 
\begin{eqnarray}
\mathcal{L}({\hat T}^{\,J}_{L,k}) \;=\; -\tfrac{1}{2}\,\gamma\, L(L+1)\,
{\hat T}^{\,J}_{L,k}. 
\label{eq:4} 
\end{eqnarray}
It then follows that the solution to (\ref{eq:1}) is 
\begin{eqnarray}
{\hat\rho}_t = \sum_{L=0}^{2J}\sum_{k=-L}^{L} 
{\re}^{-\frac{1}{2}\gamma L(L+1)t}\;\rho_{Lk}\,\hat T^{\,J}_{L,k}\, , 
\label{eq:5}
\end{eqnarray}
where $\rho_{Lk}$ are the components of the initial density matrix defined in 
(\ref{eq:3}). We see therefore that the effect of phase-space decoherence is to 
exponentially damp the elements of the density matrix. In particular, in the large 
time limit, only the $L=k=0$ term in the sum (\ref{eq:5}) survives, and from 
${\hat T}^{\,J}_{0,0} = \frac{1}{\sqrt{2J+1}}\mathds{1}$ we thus see that 
${\hat\rho}_t$ converges asymptotically to the state of complete ignorance. 

\section{Coherent-state POVM measurements}
\label{sec:3}
 
As an alternative to the Lindblad dynamics, one can model the effects of 
phase-space measurements with an application of a sequence of coherent-state 
POVMs \cite{WeigertBusch,Appleby}. Under such a POVM, if we start from an 
arbitrary initial state ${\hat\rho}$ of a spin-$J$ system and 
perform a measurement of the phase-space point of the system on the sphere, 
then the output state when the measurement outcome is recorded is given by the 
spin-$J$ coherent state centred at that point. 
In terms of the complex parameterisation 
$z=\tan{(\frac{\theta}{2})}\exp{(\ri\phi)}$ for points on phase space, where 
$\theta$ and $\phi$ are the usual spherical coordinates, the coherent state 
is given by 
\cite{Radcliffe,Arecchi,Perelomov1986,Zhang1990,Gazeau}: 
\begin{eqnarray}
\nonumber |z\rangle &=& (1 + |z|^2)^{-J} \re^{z\hat J_-}|J,J\rangle\\
&=& (1 + |z|^2)^{-J}\sum_{k=-J}^{J} \sqrt{\binom{2J}{J+k}} \, z^{J-k} \,
|J,k\rangle\, .
\label{eq:12}
\end{eqnarray}
Here $\{|J,k\rangle \}_{k=-J}^{J}$ is the eigenbasis of $\hat{J}_z$. 
The coherent states $|z\rangle$ form an overcomplete basis of 
the Hilbert space, with the resolution of the identity 
\begin{eqnarray}
\int |z\rangle \langle z| \rd\mu_{\theta,\phi}^J=\mathds{1}, 
\end{eqnarray}
where 
\begin{eqnarray}
\rd\mu^J_{\theta,\phi} = \frac{2J+1}{4\pi} \, \sin{\theta}\,\rd\theta\,\rd\phi  
\end{eqnarray}
denotes the spherical measure over the spin phase space \cite{BG}. 
In this measurement the probability of detecting a phase-space event 
\cite{BH0,BH2} in a region $A$ of the phase space is given by
\begin{eqnarray}
\mathbb{P}(z \in A) = \int_A \langle z|\hat \rho|z\rangle \,\rd \mu^J_{\theta,\phi}.
\end{eqnarray}
If the phase-space coordinates (i.e. the direction of the spin) are monitored by the 
environment, then no outcome is recorded, and the state of the system 
decoheres. In particular, writing ${\hat\rho}^{(1)}$ for the outcome state of the 
system after a single POVM measurement, we have 
\begin{eqnarray}
{\hat\rho}^{(1)} =  \int \langle z|\hat\rho|z\rangle|z\rangle\langle z| \, \rd \mu^J_{\theta,\phi} \, . 
\label{eq:x10}
\end{eqnarray}
To compare the effect of the POVM measurement with the solution (\ref{eq:5}) to 
the Lindblad equation, we shall begin by showing that the irreducible tensor 
operators, as well as being eigenstates of the Lindblad operators, are simultaneous 
eigenstates of the POVM operation. That is, writing 
\begin{eqnarray}
\Phi[{\hat A}]=\int \langle z|{\hat A}|z\rangle |z\rangle\langle z| \, \rd \mu^J_{\theta,\phi}
\end{eqnarray}
for any operator ${\hat A}$, we have $\Phi[{\hat T}^{\,J}_{L,k}]\propto 
{\hat T}^{\,J}_{L,k}$. To show this, we expand the coherent state projector 
in terms of $({\hat T}^{\,J}_{L,k})^\dagger$ by writing 
\begin{eqnarray}
|z\rangle\langle z| = \sum_{L=0}^{2J} \sum_{k=-L}^L 
\overline{c_{L,k}(z)} \,({\hat T}^{\,J}_{L,k})^\dagger, 
\end{eqnarray}
where $c_{L,k}(z) = \langle z| ({\hat T}^{\,J}_{L,k})^\dagger|z\rangle$. 
The expansion coefficients admit the representation \cite{Varshalovich, Perelomov1986,Arecchi,KlimovChumakov2009}
\begin{eqnarray}
c_{L,k}(z) = \frac{1}{\sqrt{2J+1}}\,
\left[\frac{\binom{2J}{L}}{\binom{2J+L+1}{L}}\right]^{\frac{1}{2}}
\, Y^{k}_{L}(\theta,\phi)
\end{eqnarray}
where $Y^{k}_{L}(\theta, \phi)$ denotes the spherical harmonics
\cite{Biedenharn,Varshalovich,Edmonds,Rose,Cohen}, satisfying
the orthonormality condition\footnote{We use the convention $Y^0_0=1$, 
so that the spherical harmonics are orthonormal with respect to the uniform 
probability measure $\rd\mu^0_{\theta,\phi}=(4\pi)^{-1}\sin\theta\,\rd\theta\,\rd\phi$.
This differs from the Condon--Shortley convention by a factor of $\sqrt{4\pi}$: 
$Y^m_L = \sqrt{4\pi}\,Y^{m,\mathrm{CS}}_L$, 
or equivalently $Y^m_L = \sqrt{2L+1}\,Y^{m,\mathrm{Racah}}_L$,
where $Y^{m,\mathrm{Racah}}_L = (-1)^m\sqrt{(L-m)!/(L+m)!}\,P^m_L(\cos\theta)\,\re^{\ri m\phi}$
for $m\geq0$.}
\begin{eqnarray}
\int Y^{k_1}_{L_1} \overline{Y^{k_2}_{L_2}}\,\rd \mu^0_{\theta,\phi}
= \delta_{L_1L_2}\,\delta_{k_1 k_2} \, .
\label{eq:x8}
\end{eqnarray}
Hence the map 
\begin{eqnarray}
\Phi[{\hat T}^{\, J}_{L,k}]= \int |z\rangle\langle z| {\hat T}^{\,J}_{L,k}
|z\rangle\langle z| {\rd}\mu^J_{\theta,\phi}  
\end{eqnarray}
on ${\hat T}^{\,J}_{L,k}$ involves integration over a pair of spherical harmonics, 
which on account of the orthogonality relation (\ref{eq:x8}) simplifies to give 
\begin{eqnarray}
\Phi[{\hat T}^{\,J}_{L,k}] = \frac{\binom{2J}{L}}{\binom{2J+L+1}{L}}\; 
{\hat T}^{\,J}_{L,k}. 
\label{eq:16} 
\end{eqnarray}
Thus the operators ${\hat T}^{\,J}_{L,k}$ form eigenstates of the POVM map 
$\Phi$ with the eigenvalues given by the ratio of binomial coefficients in 
(\ref{eq:16}). 

Equipped with this result, we are able to determine the effect of a phase-space 
POVM measurement on an arbitrary initial state ${\hat\rho}$. Specifically, expanding 
${\hat\rho}$ in the form (\ref{eq:3}) we find at once that after the application of 
a single measurement the expansion coefficients transform according to 
\begin{eqnarray}
\rho_{Lk} \to  \left[\binom{2J}{L}\Big/\binom{2J+L+1}{L}\right]\rho_{Lk} . 
\label{eq:17} 
\end{eqnarray}
It also follows, more generally, that an application of $n$ successive 
measurements will transform the state into 
\begin{eqnarray}
\hat{\rho}^{(n)} = \sum_{L=0}^{2J}\sum_{k=-L}^{L} 
\left[\frac{\binom{2J}{L}}{\binom{2J+L+1}{L}}
\right]^{n} \rho_{Lk} \, {\hat T}^{\,J}_{L, k} \, . 
\label{eq:18} 
\end{eqnarray}
Analogous to the solution (\ref{eq:5}) to the Lindblad dynamics, the off-diagonal 
elements of the density matrix are exponentially suppressed, and the state 
eventually decays into the stable state $\rho_{00} 
{\hat T}^{\,J}_{0,0} = \frac{1}{2J+1}\mathds{1}$, 
the state of complete ignorance. 

\section{Comparison of the decoherence rates} 
\label{sec:4} 

Although the two models of phase-space decoherence give rise to the same 
asymptotic results, the decoherence rates are not identical. Specifically, the 
decay rates for the Lindblad process are given by
\begin{eqnarray}
\Gamma^{\rm(Lind)}_{L}=\frac{1}{2} \gamma L(L+1),
\label{eq:x30} 
\end{eqnarray}
while the corresponding decay rates of the sequence of POVM measurements, 
when we view the iteration count $n$ as representing time, are given by
\begin{eqnarray}
\Gamma^{\rm(POVM)}_{L}= 
- \log\left(\frac{{2J \choose L}}{{2J+L+1 \choose L}}\right)\, . 
\label{eq:19}
\end{eqnarray}
Both decay rates are independent of the value of $k$. 
While in both cases $\Gamma_{0}=0$, the decay rates for the other $L\neq0$ 
components differ in general, except in the case $J=1/2$. For $J=1/2$ there is 
only one nonzero decay rate in both cases, and they agree under the choice 
$\gamma=\log3$. In this case the solution to the Lindblad equation at $t=n$ 
agrees with the outcome state resulting from $n$ iterated POVM measurements. 

The behaviours of the two models become analogous also for large spins in 
the following sense. When $L$ is held fixed, for large spin $J$ we have 
\begin{eqnarray}
\Gamma^{\rm(POVM)}_{L} &\approx& \frac{1}{2J} L(L+1) - \frac{1}{4J^2} L(L+1) 
\nonumber \\ && + \frac{1}{48J^3} L(L+1)(L^2+L+6) \, , 
\label{eq:x32} 
\end{eqnarray} 
and hence 
$\Gamma^{\rm(POVM)}_{L} \to \Gamma^{\rm(Lind)}_{L}$ if we let $\gamma = 
1/J$. However, if $L$ scales with $J$, and recall that $L$ ranges from $0$ to 
$2J$, then the decay rates do not agree. Nevertheless, in this case the large-$L$ 
components of the density matrix in either model will decay so fast anyhow 
that they will make little practical difference. In this sense the two models behave 
similarly for large spins. For intermediate 
spins, the two decoherence models are not equivalent, although their dynamical 
characteristics agree qualitatively, and in the asymptotic 
limits $t \to \infty$ and $n\to\infty$ the off-diagonal terms are suppressed, 
while the diagonal elements approach a uniform distribution, independently 
of the choice of the basis in which the density matrix is represented. 

\begin{figure}[t] 
    \centering
    \includegraphics[width=0.48\textwidth]{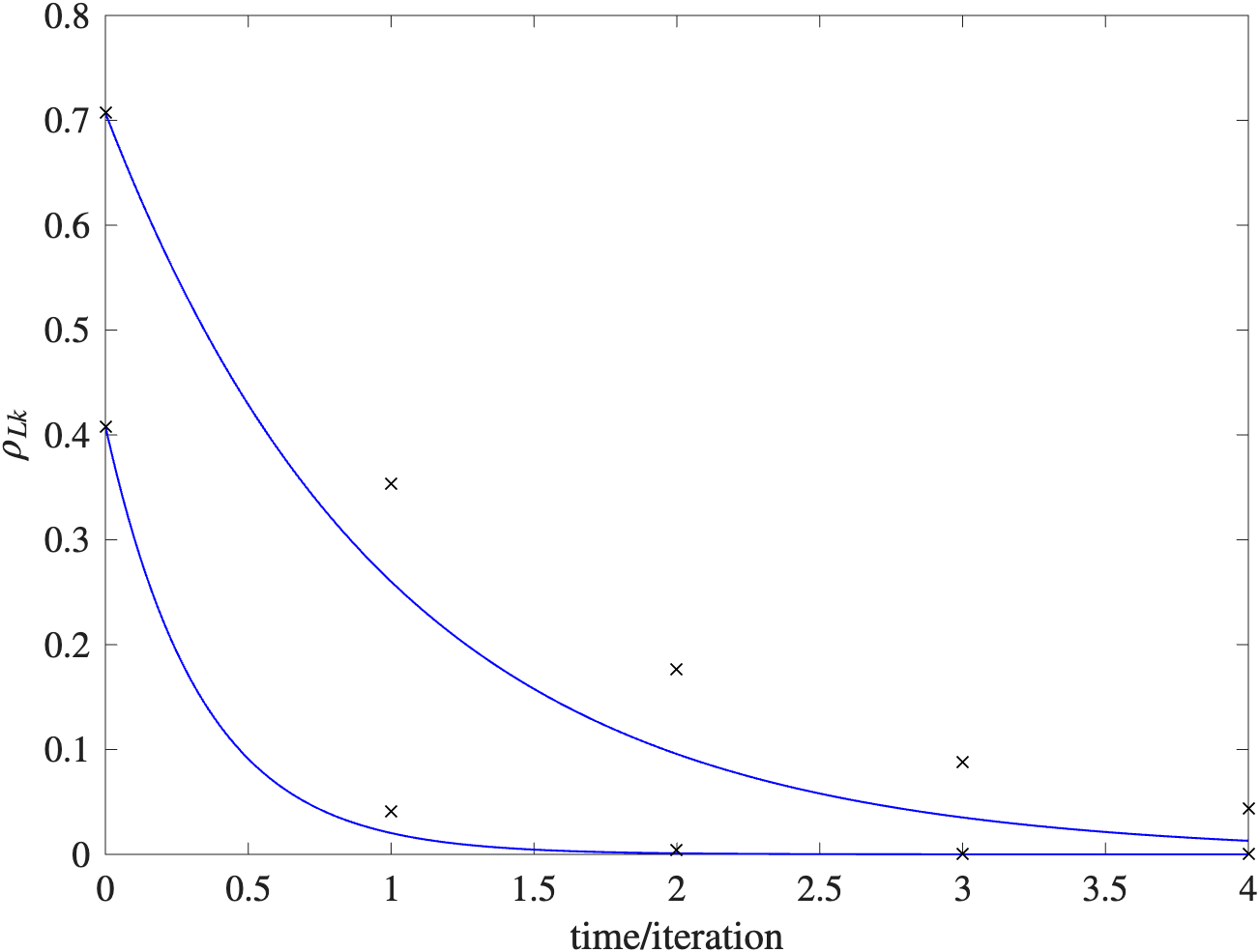} 
             \includegraphics[width=0.48\textwidth]{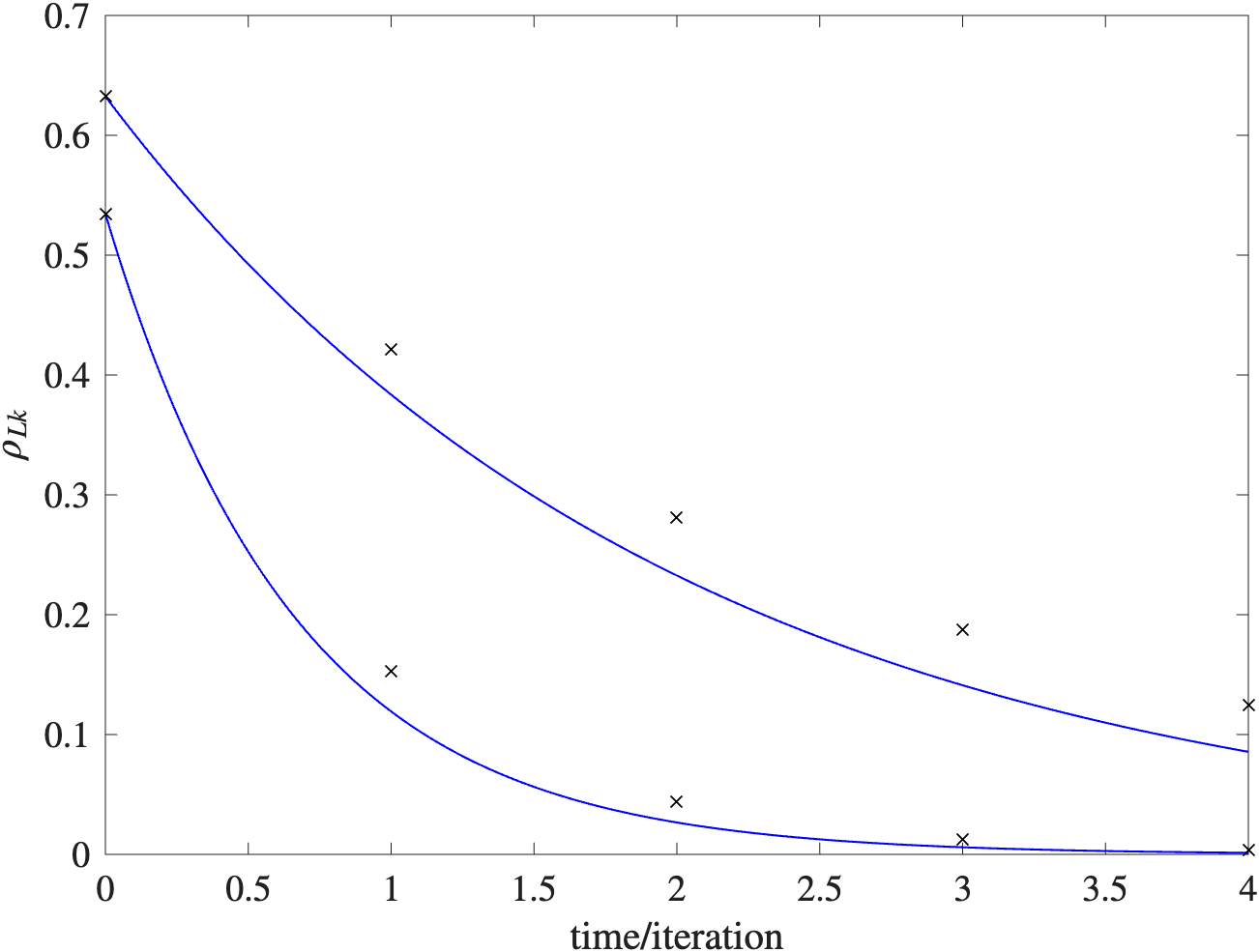}
    \caption{Decoherence effect for $J=1$ (left) and $J=5$ (right). 
The components $\rho_{20}$ and $\rho_{10}$ of the density matrix, 
prepared initially in a coherent state, as functions of time/iteration 
number for the Lindblad dynamics (solid blue lines, with $\gamma=1/J$) 
and repeated POVM measurements (black crosses) are shown here. 
The impact of decoherence in terms of the decay of the elements of the 
density matrix is less pronounced with coherent states having larger 
spins.}
    \label{fig1}
\end{figure}

The existence of the two distinct yet plausible models for the decoherence 
arising from an isotropic environmental monitoring of the spin raises the 
question of whether a physical environment might preferentially follow one 
over the other (or follow a third option). 
To distinguish the dynamics identified here in a possible experiment, 
one could examine the relative decay of different multipole moments.
Recall that we are comparing decoherence models with two decay rates 
(\ref{eq:x30}) and (\ref{eq:19}). For 
$L=1,2$ (dipole and quadrupole sectors), the corresponding decay rates do 
not coincide. 
In Figure \ref{fig1} we display examples of both the POVM and the Lindblad 
dynamics, for an initial coherent state pointing towards the north pole 
(``spin up'') for $J=1$ and $J=5$, respectively, where we have set 
$\gamma=1/J$ for the parameter in the Lindblad dynamics. Specifically, the 
components $\rho_{Lk}$ of the density matrix for $L=1,2$ and $k=0$ for 
both the Lindblad evolution (solid blue lines) and iterated POVMs (denoted 
by black crosses) are shown. For both values of $J$ the decay of these 
elements is visibly different between the two models.
In general, the value of $\gamma$ is arbitrary, so it is better to consider 
ratios of decay rates. For spin-1, for example, we find 
\begin{eqnarray}
\frac{\Gamma^{\rm(Lind)}_{2}}{\Gamma^{\rm(Lind)}_{1}} = 3 
\quad {\rm and} \quad 
\frac{\Gamma^{\rm(POVM)}_{2}}{\Gamma^{\rm(POVM)}_{1}} = 
\frac{\log 10}{\log 2} \approx 3.32\, .
\end{eqnarray}
These dimensionless numbers provide a subtle discrimination between 
the two models.

These decay rates can be accessed from monitoring the so-called orientation 
sector $\rho_{1k}$ and the alignment sector $\rho_{2k}$. These are the 
coefficients of the spin-$1$ density matrix in the irreducible tensor basis 
\cite{Budker2002,Rochester2001,Geremia2005,Sewell2012,
Rochester2012, Stahovich2024}). 
In an experimental setting where it is expected that an environment monitors 
the spin without isolating a preferred orientation (for a generic environment, a 
symmetry argument requires that no orientation should be preferred), one 
could empirically determine which model is more representative by measuring 
$\rho_{1k}(t)$ and $\rho_{2k'}(t)$ after some time $t$, having prepared the 
initial state of the system in some $\rho_{Lk}(0)$, and forming the empirical ratio 
\begin{eqnarray}
R = \frac{\log \big(|\rho_{2k'}(t)|/|\rho_{2k'}(0)|\big)}
         {\log \big(|\rho_{1k}(t)|/|\rho_{1k}(0)|\big)}
\end{eqnarray}
for any $k,k'$, which theoretically will be time-independent. 
The value of $R$ then 
determines the ratio of the decay rates. 
While both models agree qualitatively in their approach 
of the state of complete ignorance, the specific value of $R$ acts as a fingerprint of the 
underlying processes described by a continuous Lindblad monitoring or discrete, iterated POVM measurements.

\section{Representations in terms of phase-space quasiprobability functions} 
\label{sec:5} 

\begin{figure*}[t] 
    \centering
     \includegraphics[width=0.32\textwidth]{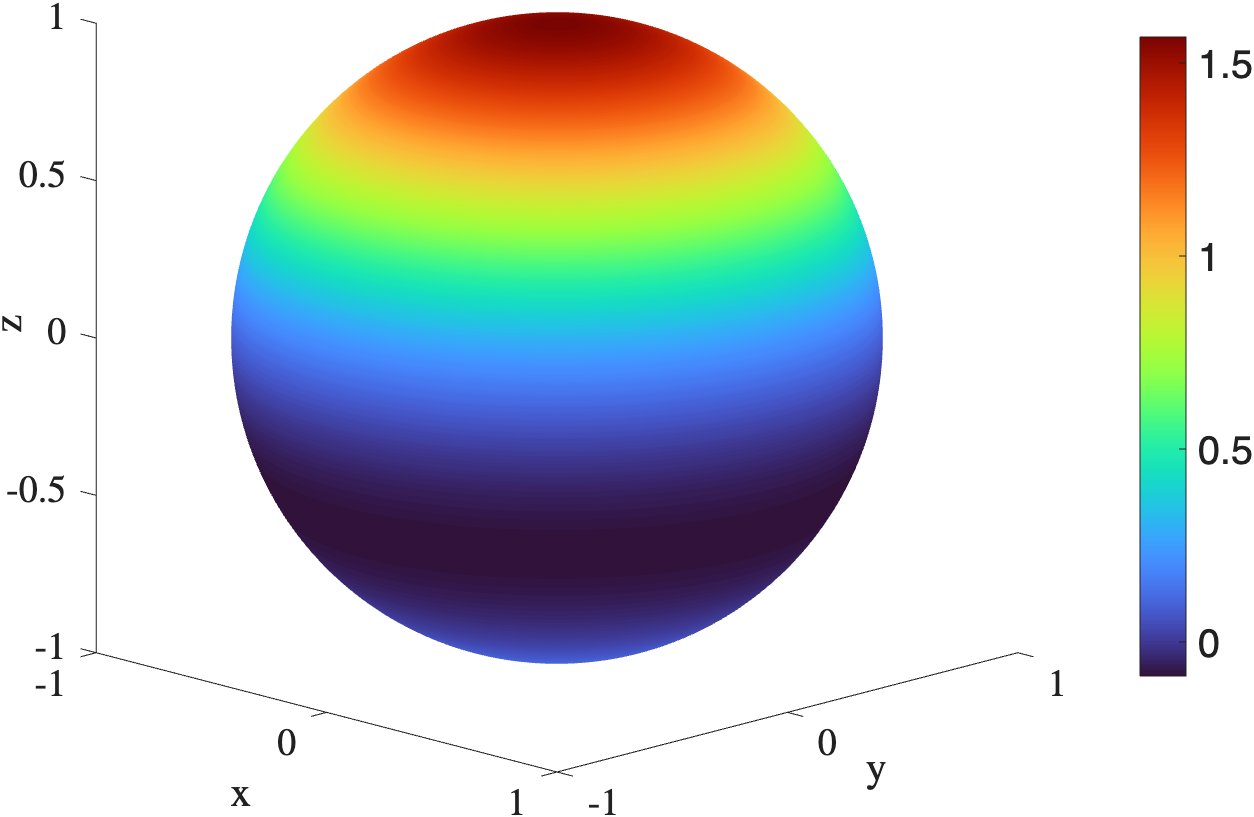}
     \includegraphics[width=0.32\textwidth]{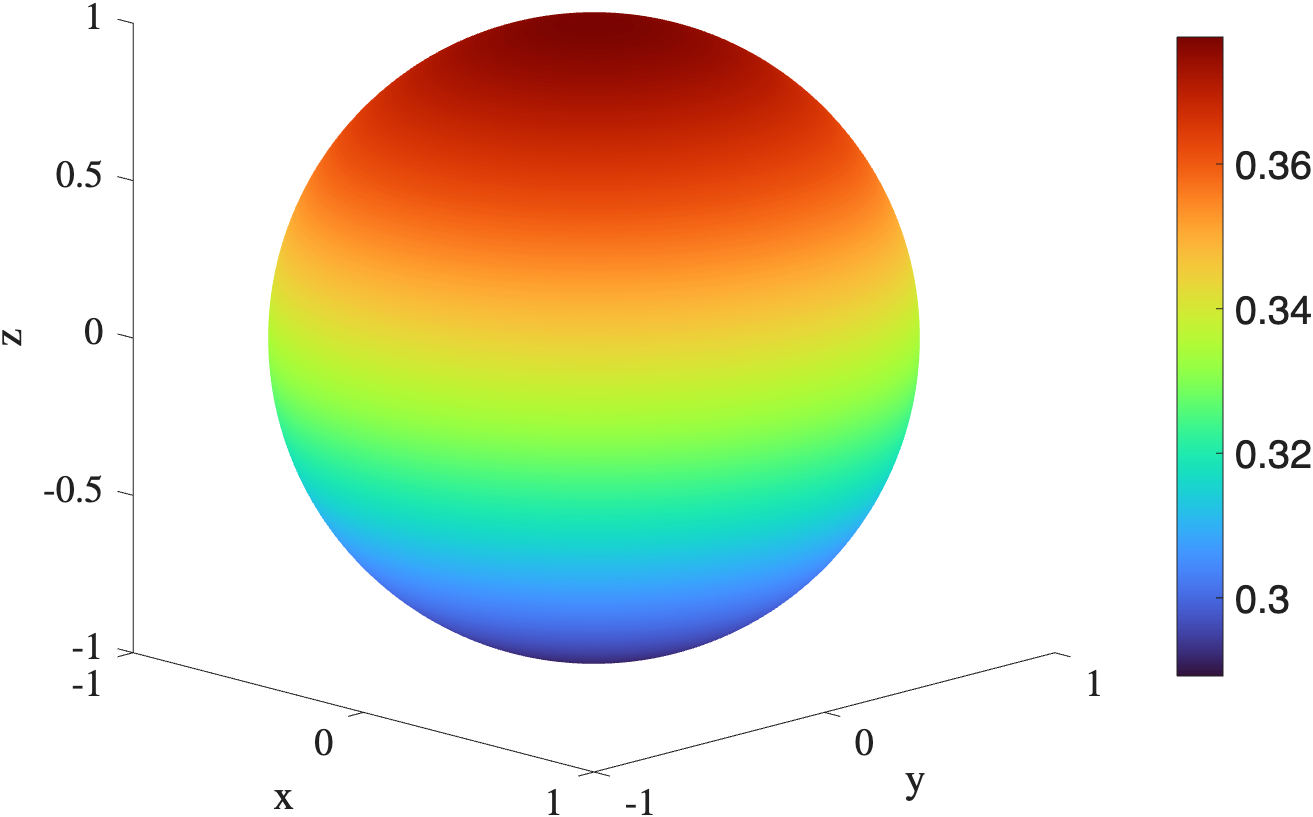} 
         \includegraphics[width=0.32\textwidth]{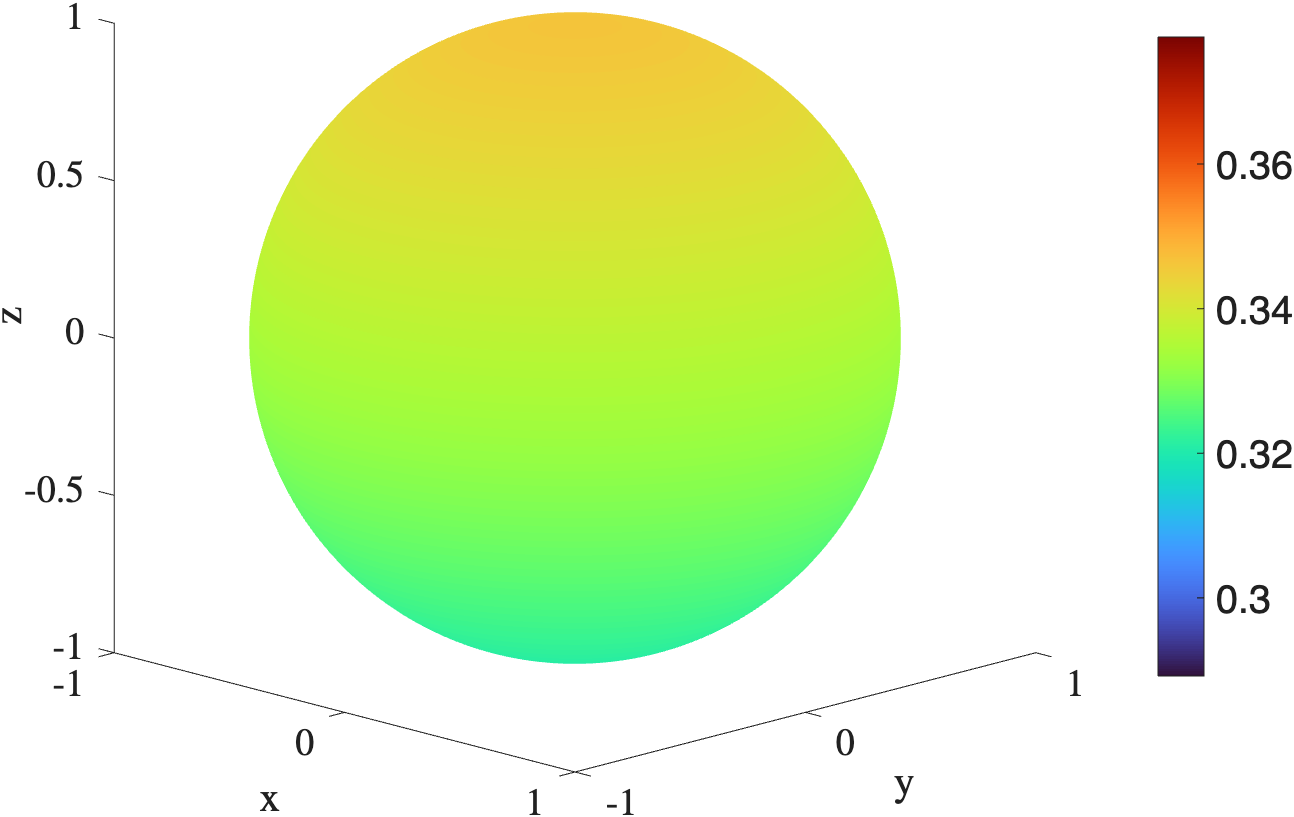} \\
     \includegraphics[width=0.32\textwidth]{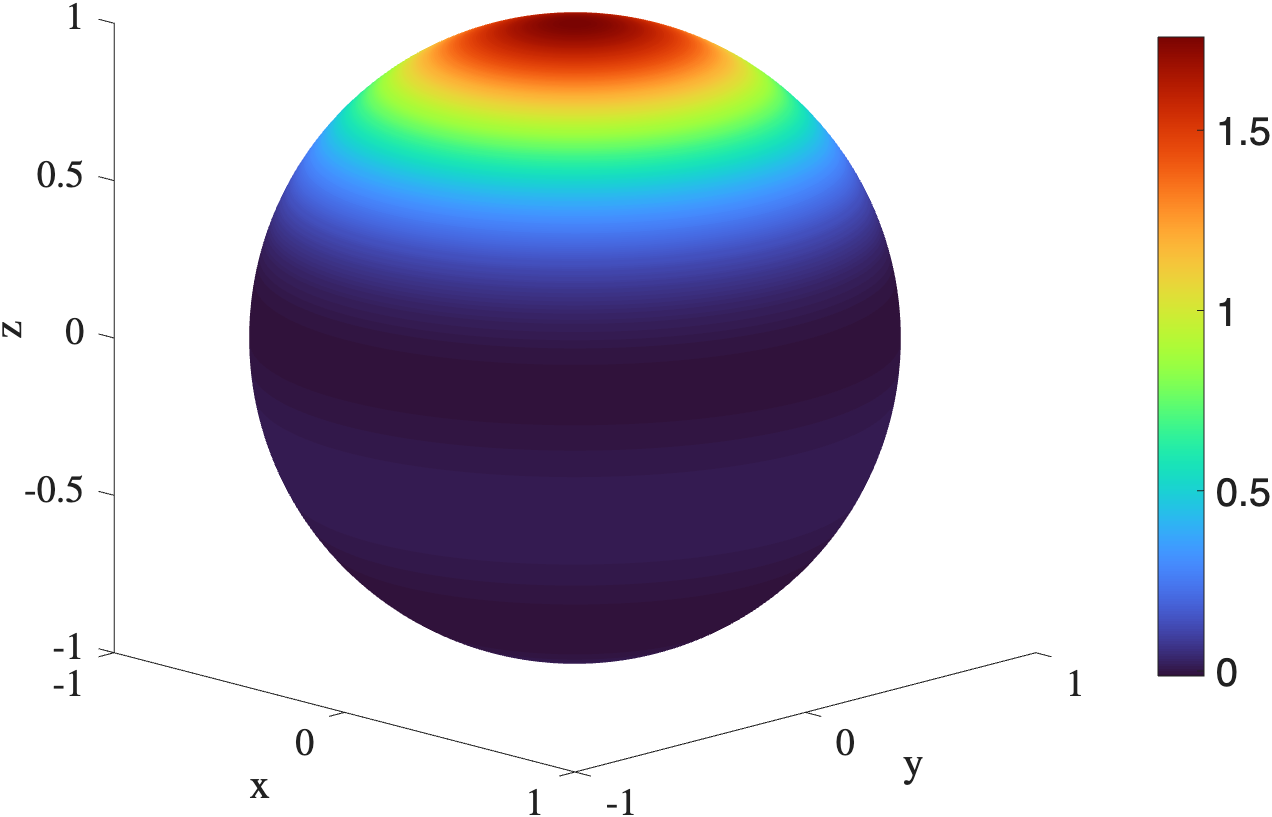}
     \includegraphics[width=0.32\textwidth]{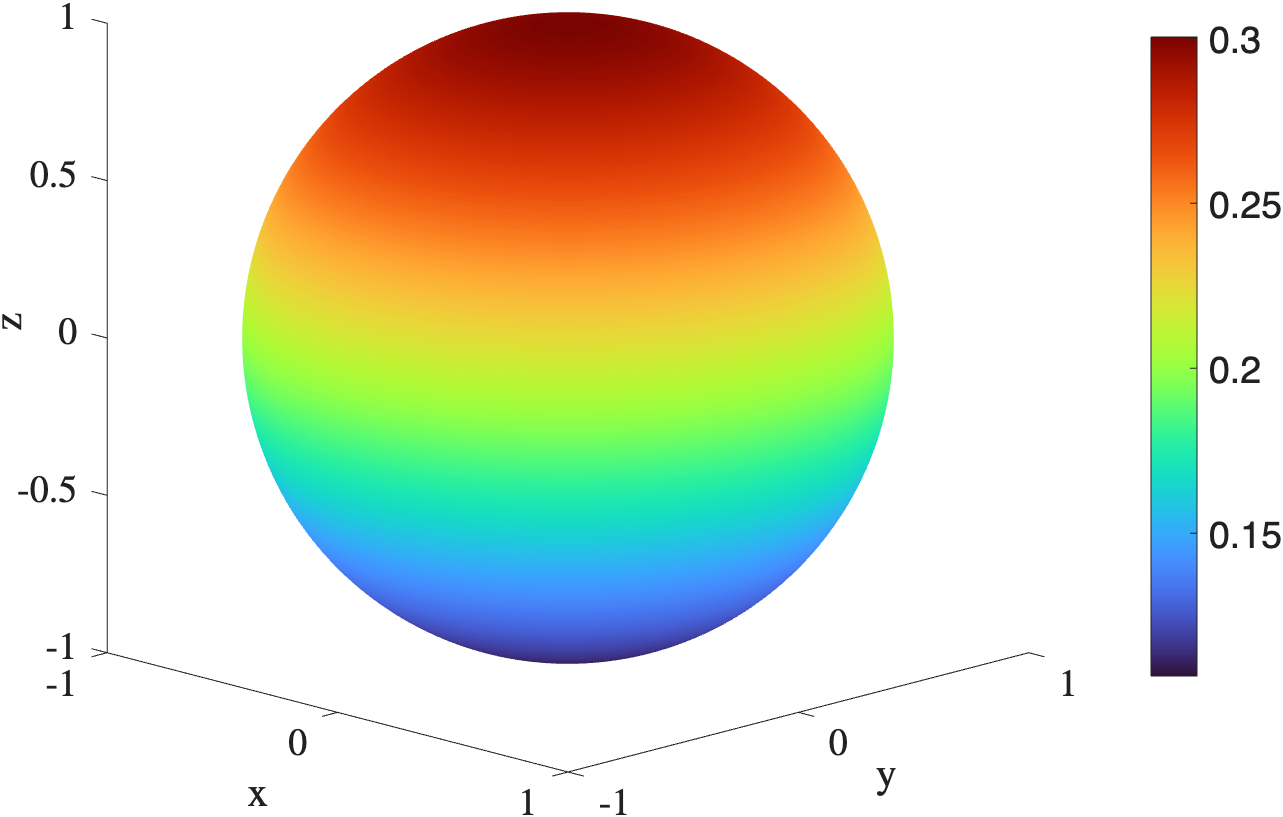} 
         \includegraphics[width=0.32\textwidth]{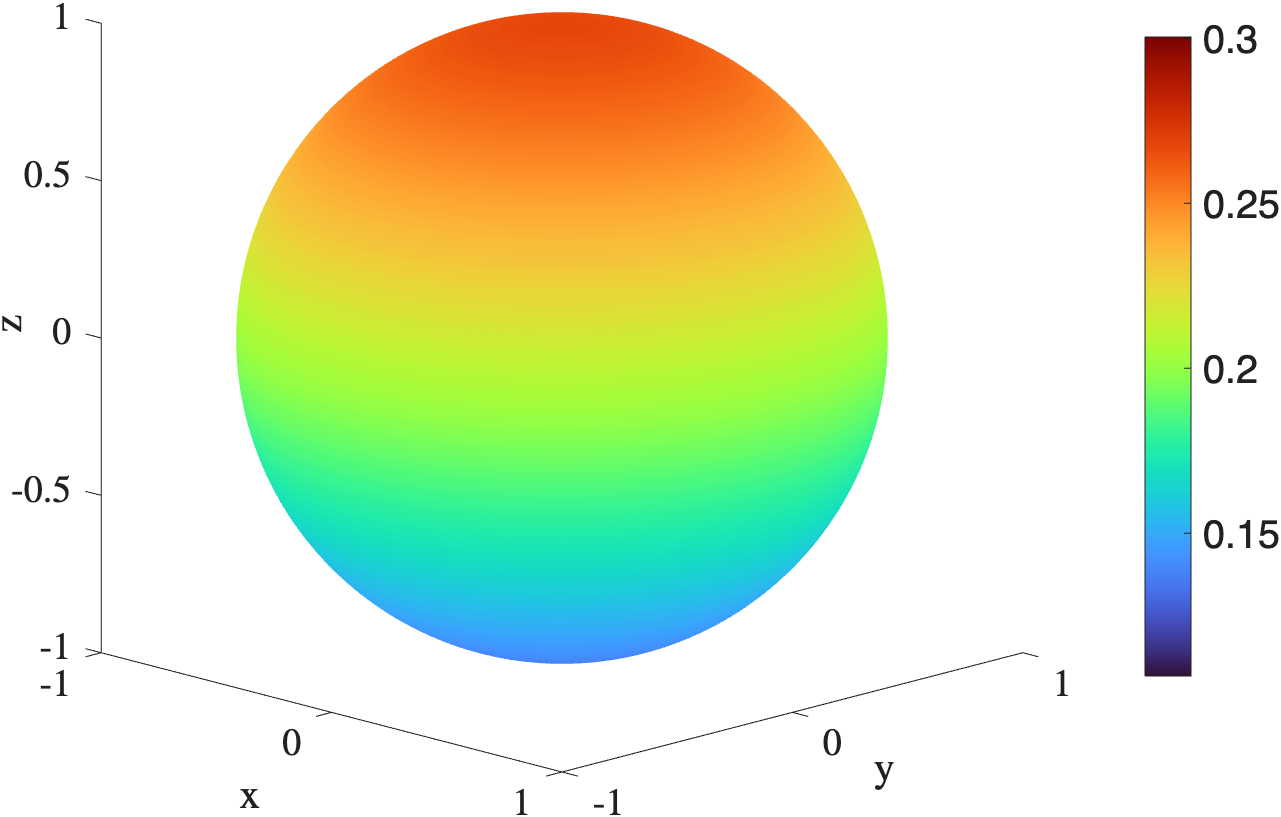}
    \caption{Comparison of the decoherence models for an initial spin coherent 
    state for $J=1$ (top) and $J=5$ (bottom). The left column shows the Wigner 
    functions of initial coherent states with $J=1$ and $J=5$. The middle column 
    shows the transformed Wigner functions after four iterations of POVM 
    measurements. The right column shows the Wigner functions as the 
    solution to the Lindblad evolution at time $t=4$ (with $\gamma=1/J$). The 
    results illustrate the fact that features of a coherent state with higher spins 
    are more robust against decoherence.}
    \label{fig2}
\end{figure*}

To gain further insights into the decoherence effects of the Lindblad dynamics 
(\ref{eq:1}) and the POVM map (\ref{eq:x10}), we consider the 
corresponding dynamical equation and the transformation rule 
satisfied by phase-space quasiprobability 
functions. To this end we shall be working with a parametric family 
of quasiprobability distributions for $\mathsf{SU}(2)$ systems 
\cite{Stratanovich, Klimov2002, Klimov20022, Klimov2017} defined by 
\begin{eqnarray}
F^{\sigma}(\theta, \phi) = {\rm tr}\left(\hat{\rho} \, \hat{w}^{\sigma}(\theta,\phi)\right) 
\label{eq:6} 
\end{eqnarray}
on the spherical phase space associated with a given quantum state ${\hat\rho}$.  
Here the Stratonovich-Weyl kernel  
\begin{eqnarray}
\hat{w}^{\sigma}(\theta, \phi) = a_J 
\sum_{L=0}^{2J} \sum_{k=-L}^{L}
\left[\frac{\binom{2J}{L}}{\binom{2J+L+1}{L}} \right]^{-\frac{\sigma}{2}}
Y^k_{L}(\theta, \phi) \, {\hat T}^{\,J}_{L,k} \nonumber \\
\label{eq:7}
\end{eqnarray}
satisfies the normalisation 
$\int \hat{w}^{\sigma}(\theta, \phi) \, {\rd}\mu_{\theta,\phi}^{J} = \mathds{1}$ 
and the unit-trace condition ${\rm tr}\!\left(\hat{w}^{\sigma}(\theta, \phi)\right)=
1$. Here we have defined $a_J=(2J+1)^{-1/2}$. 

The normalisation of the kernel $\hat{w}^{\sigma}(\theta, \phi)$ implies that 
$\int F^{\sigma}(\theta, \phi) \, {\rd}\mu_{\theta,\phi}^{J} = 1$, but in general  
the positivity condition $F^\sigma(\theta,\phi)\geq0$ is not satisfied. 
The family $\{F^{\sigma}\}$ includes, in particular, the Husimi $Q$ function 
($\sigma=-1$), which is manifestly positive \cite{Husimi}, the Wigner-Stratonovic 
$W$ function 
($\sigma=0$)~\cite{Wigner1932}, and the Glauber-Sudarshan $P$ function 
($\sigma=+1$)~\cite{Sudarshan}, 
for $\mathsf{SU}(2)$ systems. The negativity of the 
quasiprobability distribution is progressively suppressed with decreasing $\sigma$. 

While we can directly deduce the time-dependent phase-space distribution from the 
time-dependent density matrices derived in sections \ref{sec:2} and \ref{sec:3}, it is 
instructive to consider the dynamical equations for the phase space distributions 
arising from the Lindblad dynamics (\ref{eq:1}). 
We will show that $F^{\sigma}(\theta,\phi,t)$ satisfies the heat 
equation on the sphere:
\begin{eqnarray}
\frac{\rd F^\sigma}{\rd t} =  \frac{\gamma}{2}\left(\frac{\partial^{2}}
{\partial \theta^{2}}+\cot{\theta} \frac{\partial}{\partial \theta} 
+ \frac{1}{\sin^{2}{\theta}} 
\frac{\partial^{2} }{\partial \phi^{2}}\right) F^\sigma.
\label{eq:9} 
\end{eqnarray}
To derive (\ref{eq:9}), we use the self-adjointness of $\mathcal{L}$ 
with respect to the Hilbert-Schmidt inner product to find 
\begin{eqnarray}
\partial_t F^{\sigma} =
{\rm tr}\!\left(\mathcal{L}(\hat{\rho})\,\hat{w}^{\sigma}\right)
= {\rm tr}\!\left(\hat{\rho}\,\mathcal{L}(\hat{w}^{\sigma})\right).
\end{eqnarray}
To evaluate $\mathcal{L}(\hat{w}^{\sigma})$, note that 
$\mathcal{L}$ acts only on the operator part ${\hat T}^{\,J}_{L,m}$ 
of each term in the sum (\ref{eq:7}). Hence on account of (\ref{eq:4}) 
we obtain
\begin{eqnarray}
\mathcal{L}\!\left(\hat{w}^{\sigma}\right)
&=& a_J \sum_{L,k}
\left[\frac{\binom{2J}{L}}{\binom{2J+L+1}{L}} \right]^{-\frac{\sigma}{2}}
Y^{k}_{L} \,\mathcal{L}\!\left(\hat T^{\,J}_{L,k}\right)
\nonumber \\&& \hspace{-1.4cm}
= \frac{\gamma}{2}\,a_J \sum_{L,k}
\left[\frac{\binom{2J}{L}}{\binom{2J+L+1}{L}} \right]^{-\frac{\sigma}{2}}
\bigl(-L(L+1)\bigr)Y^{k}_{L}\,\hat T^{\,J}_{L,k}.~~~
\end{eqnarray}
On the other hand, spherical harmonics, by definition, 
satisfy $\Delta_{\mathbb{S}^2}Y_{L}^k = -L(L+1)\,Y_L^k$, 
where 
\begin{equation}
\Delta_{\mathbb{S}^2}=\frac{\partial^{2}}
{\partial \theta^{2}}+\cot{\theta} \frac{\partial}{\partial \theta} 
+ \frac{1}{\sin^{2}{\theta}} 
\frac{\partial^{2} }{\partial \phi^{2}}
\end{equation}
denotes the angular part of the Laplacian that appears in (\ref{eq:9}). 
Hence we deduce that
\begin{eqnarray}
\mathcal{L}\!\left(\hat{w}^{\sigma}(\theta,\phi)\right)
= \frac{\gamma}{2}\,\Delta_{\mathbb{S}^2}\hat{w}^{\sigma}(\theta,\phi)\, , 
\label{eq:Lw}
\end{eqnarray}
from which it follows that
\begin{eqnarray}
\partial_t F^{\sigma}
&=& \frac{\gamma}{2}\,{\rm tr}\!\left(\hat{\rho}\,
\Delta_{\mathbb{S}^2}\hat{w}^{\sigma}\right) \nonumber \\
&=& \frac{\gamma}{2}\,\Delta_{\mathbb{S}^2}\,
{\rm tr}\!\left(\hat{\rho}\,\hat{w}^{\sigma}\right)
\;=\; \frac{\gamma}{2}\,\Delta_{\mathbb{S}^2} F^{\sigma},
\end{eqnarray}
and this establishes (\ref{eq:9}).

The solution to the heat equation (\ref{eq:9}) is found by eigenfunction
expansion. Writing
\begin{eqnarray}
F^{\sigma}(\theta,\phi,t) = a_J\sum_{L,k}
g_{Lk}(t)\, \overline{Y_L^k(\theta,\phi)}
\end{eqnarray}
for some expansion coefficients $g_{Lk}$ 
and substituting this in (\ref{eq:9}) gives $\partial_t g_{Lk} =
-\tfrac{\gamma}{2}L(L+1)\,g_{Lk}$,
so each coefficient decays exponentially according to 
$g_{Lk}(t) = \re^{-\frac{\gamma}{2}L(L+1)t}\,g_{Lk}(0)$.
The initial coefficients are determined by evaluating
$F^\sigma(\theta,\phi,0)={\rm tr}(\hat\rho(0)\,\hat{w}^{\sigma})$ directly:
substituting the expansion (\ref{eq:3}) into (\ref{eq:6})--(\ref{eq:7})
and evaluating the trace using
${\rm tr}(\hat{T}^{\,J}_{L,k}\,\hat{T}^{\,J}_{L',m})
= (-1)^{k}\delta_{LL'}\delta_{k,-m}$
together with $Y_{L}^{-k}=(-1)^{k}\overline{Y_{L}^{k}}$ gives
$g_{Lk}(0)=\bigl(\binom{2J}{L}/\binom{2J+L+1}{L}\bigr)^{-\sigma/2}\rho_{Lk}(0)$,
and hence we obtain the explicit solution
\begin{eqnarray}
F^{\sigma}(\theta,\phi,t)  &=& \nonumber \\ && \hspace{-2.1cm}
a_J \sum_{L,k}{\re}^{-\frac{\gamma}{2} L(L+1)t}\left(
\frac{{2J \choose L}}{{2J+L+1 \choose L}}\right)^{-\frac{\sigma}{2}}
\rho_{Lk}(0) \, \overline{Y^{k}_{L}(\theta,\phi)}.~~~~~
\label{eq:x44}
\end{eqnarray}
To express the solution for the quasidistribution function in terms of its initial 
condition, we use the orthogonality of spherical harmonics to deduce that 
\begin{eqnarray}
&& a_J \left(\frac{{2J \choose L}}{{2J+L+1 \choose L}}\right)^{-\frac{\sigma}{2}} 
\rho_{Lk}(0) \nonumber \\ && \qquad 
 = 
\int Y^{m}_{l}(\theta, \phi)\, F^{\sigma}(\theta, \phi, 0) \, {\rm d}\mu^{0}_{\theta,\phi} \, .
\end{eqnarray}
Substituting this in the right side of (\ref{eq:x44}) we obtain 
\begin{eqnarray}
F^{\sigma}(\theta, \phi, t)  &=& \sum_{L,k} {\re}^{-\frac{\gamma}{2}L(L+1)t}
\overline{Y^{k}_{L}(\theta, \phi)} \nonumber \\ &&
\int Y^{k}_{L}(\theta',\phi')\, F^{\sigma}(\theta',\phi',0) \,
{\rd}\mu^{0}_{\theta', \phi'}.
\end{eqnarray}
Summing over $k$, and using the addition theorem for spherical harmonics, 
finally gives the expression 
\begin{eqnarray}
F^{\sigma}(\theta,\phi,t) &=& \sum_{L=0}^{2J} {\re}^{-\frac{\gamma}{2}L(L+1)t} 
\nonumber \\ && 
\int P_{L}(\cos{\eta})\, F^{\sigma}(\theta',\phi',0) \, {\rd}\mu^{L}_{\theta',\phi'} \, , 
\label{eq:x47} 
\end{eqnarray}
where $\cos{\eta}=\cos\theta\cos\theta'
+\sin\theta\sin\theta'\cos(\phi-\phi')$ and 
\begin{eqnarray}
P_{L}\left(\cos \eta\right) = 
\sum_{k=-L}^{L} 
\frac{Y^{k}_{L}(\theta,\phi) \overline{Y^{k}_{L}(\theta',\phi')}}{2L+1} 
\label{eq:11} 
\end{eqnarray}
is the $L^{\text{th}}$ zonal harmonic \cite{Edmonds}. 

The phase-space representation of the Lindblad equation thus shows that the 
dynamics give rise to a heat equation associated with a Brownian motion on 
phase space, whose solution asymptotically converges to the uniform distribution 
$F^\sigma(\theta,\phi,t) \to (2J+1)^{-1}$ for all $\sigma$. This follows from 
(\ref{eq:x47}), which shows that in the large time limit the only surviving term in the 
sum is the one for which $L=0$, but we have $P_0=1$ and the normalisation 
condition $\int F^\sigma(\theta,\phi)\rd\mu_{\theta, \phi}^L=(2L+1)/(2J+1)$. 
Thus, irrespective of the value of $\sigma$, all
Stratonovich-Weyl phase-space quasidistributions $F^{\sigma}$ obey the same
heat equation on the sphere.

\begin{figure*}[t] 
    \centering
     \includegraphics[width=0.32\textwidth]{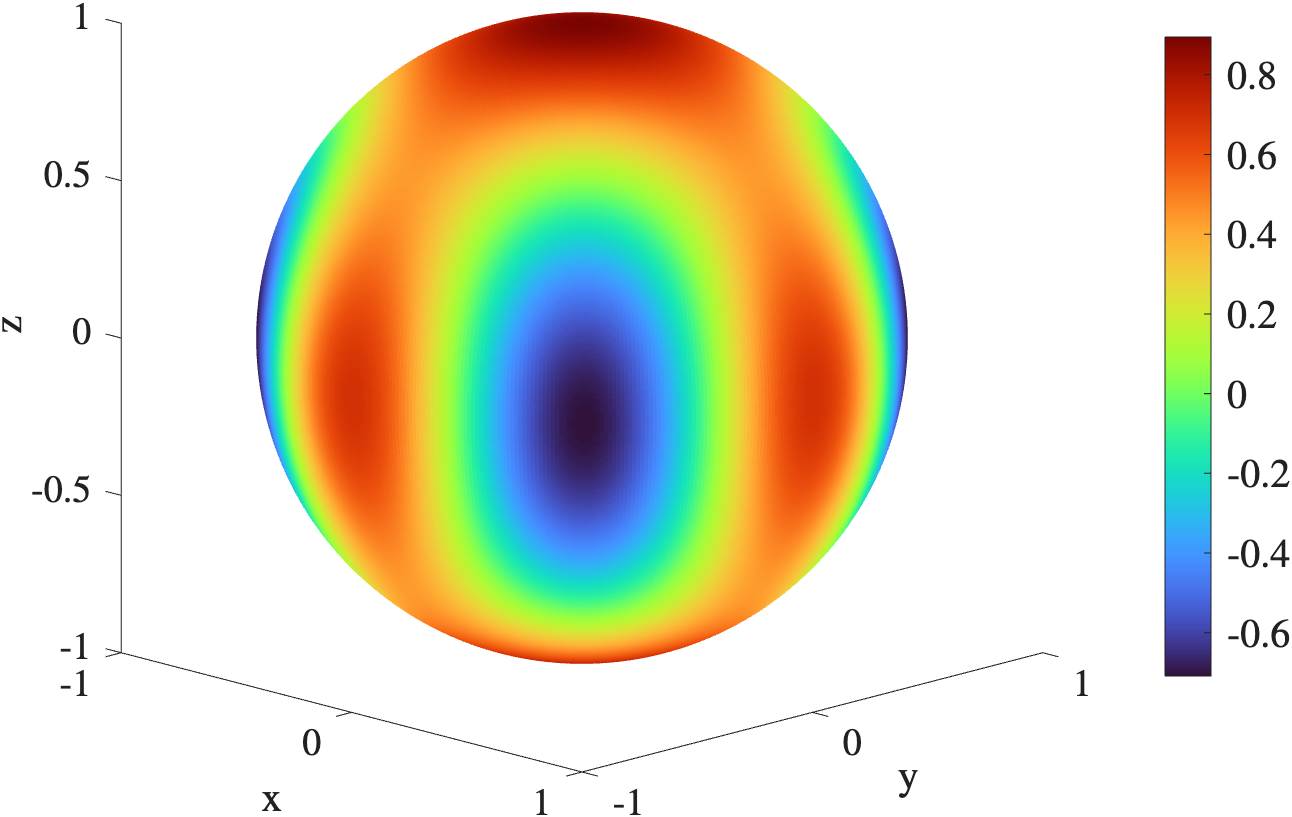}
     \includegraphics[width=0.32\textwidth]{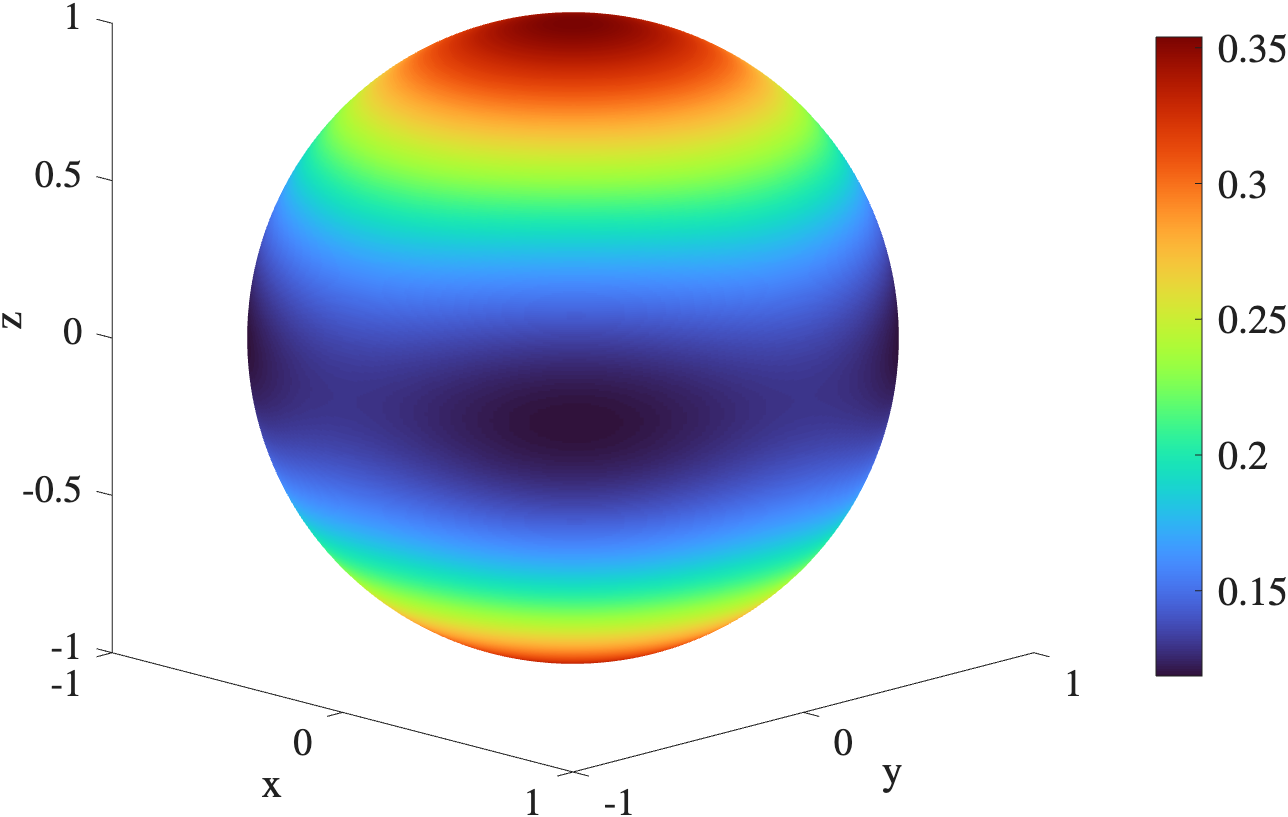} 
              \includegraphics[width=0.32\textwidth]{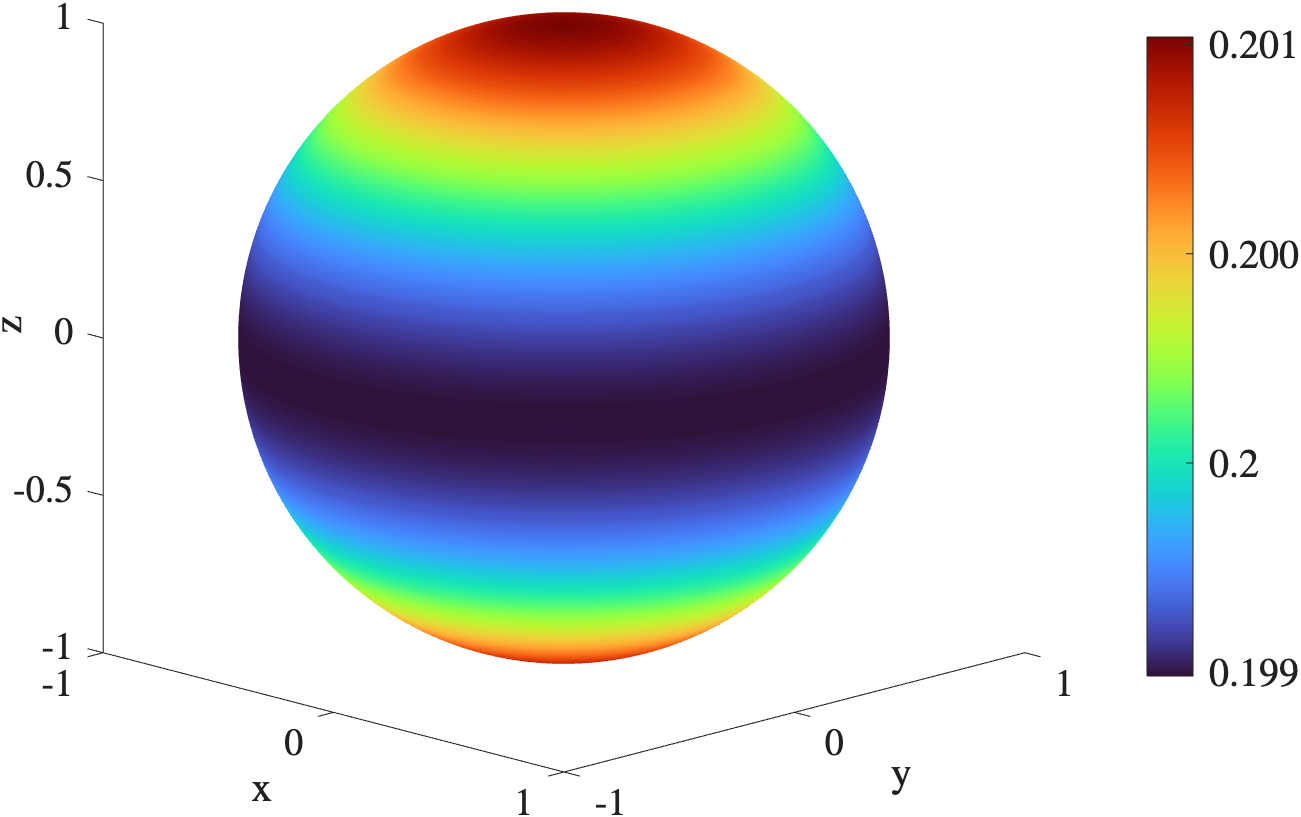} \\
     \includegraphics[width=0.32\textwidth]{Wigner_t0_NOON_J2-2.png}
     \includegraphics[width=0.32\textwidth]{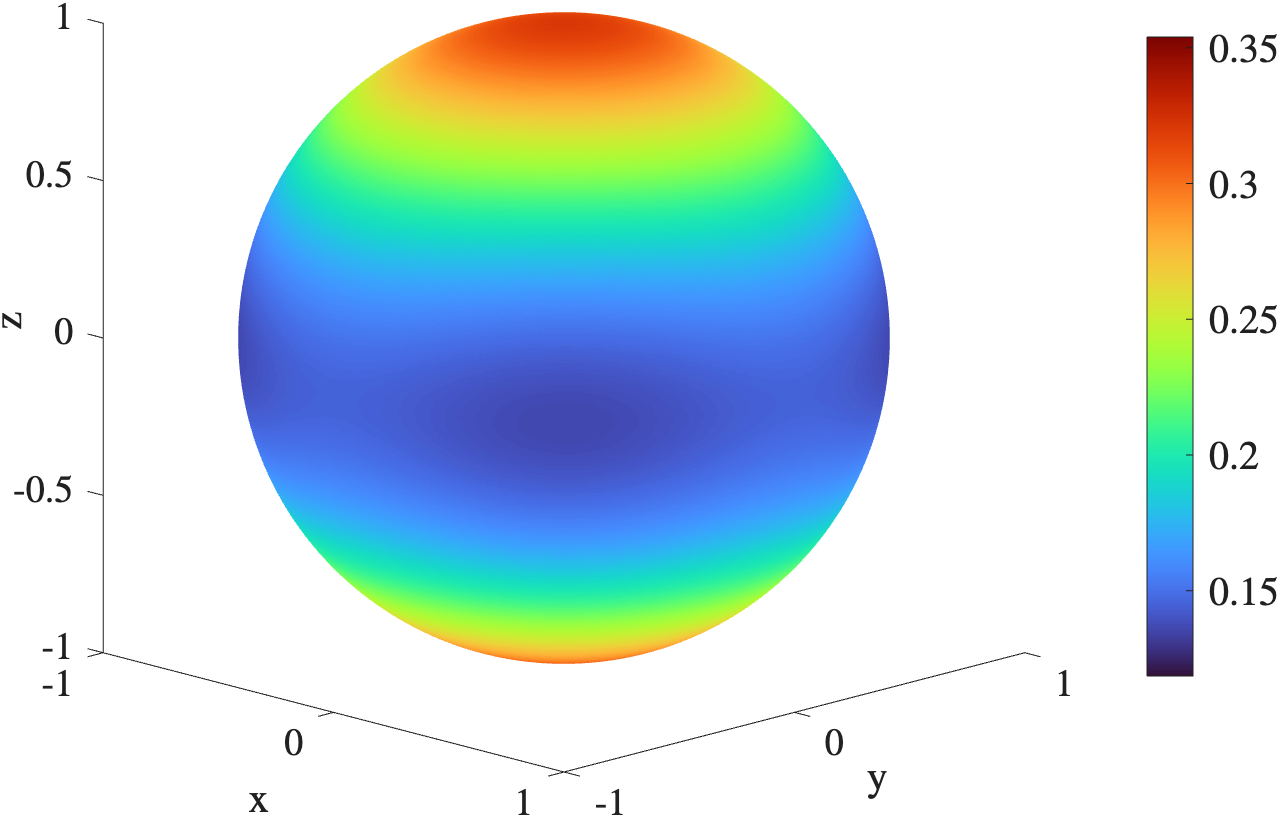} 
              \includegraphics[width=0.32\textwidth]{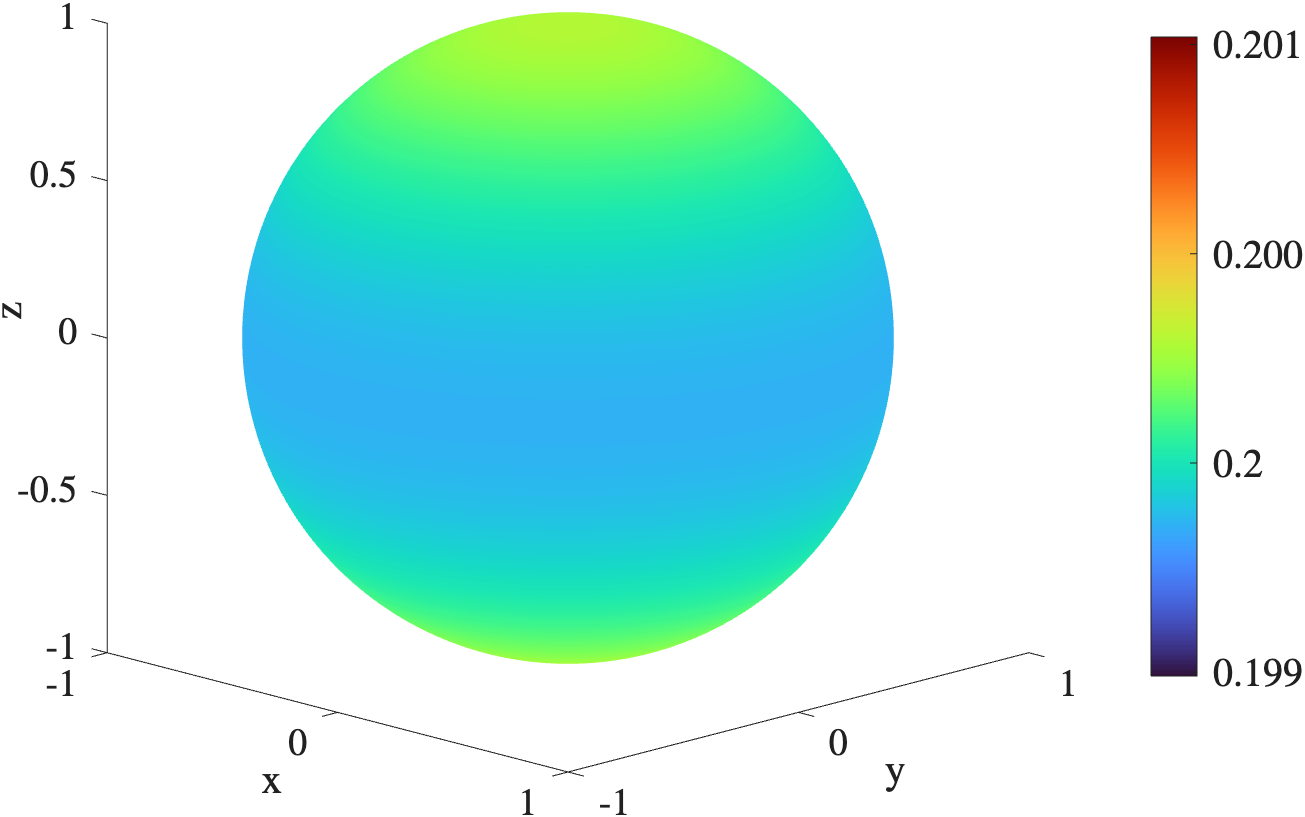} 
    \caption{Decoherence of the state $|\psi\rangle=\frac{1}{\sqrt{2}}(|m = J\rangle
    +|m = -J\rangle)$ with $J=2$. The top raw represents the effect of POVM 
    measurements 
    after one (middle) and five (right) iterations. The bottom raw shows the 
    effect of Lindbladian dynamics at times $t=1$ and $t=5$ (with $\gamma=1/J$). 
    The approach to uniform distribution resulting from decoherence is faster for 
    the Lindblad dynamics in this example. 
    }
    \label{fig3}
\end{figure*}

Next, to gain insights into the phase-space characteristics of the POVM 
measurements, consider their impacts on the phase-space quasidistribution 
function $F^{\sigma}$. For the impact of the coherent-state POVM, 
the fact that the eigenvalues of $\Phi$ appear in the expression (\ref{eq:7}) of 
the Stratonovic-Weyl kernel shows at once that the action of each measurement 
is to lower the value of $\sigma$ by~$2$. This follows from the identity
that ${\rm tr}( {\Phi}[{\hat\rho}]\hat{w}^{\sigma})={\rm tr}( {\hat\rho} \,
{\Phi}[\hat{w}^{\sigma}])$, on account of the self-adjointness of the 
phase-space measurement operation. Hence after performing a single 
phase-space measurement, the quasidistribution function transforms according 
to $F^{\sigma} \mapsto F^{\sigma-2}$. More generally, writing 
$F^{\sigma}(\theta,\phi,n)$ for the quasidistribution function resulting from 
$n$ successive POVM measurements, we have 
\begin{eqnarray} 
F^{\sigma}(\theta,\phi,n) = F^{\sigma-2n}(\theta,\phi,0) \, . 
\label{eq:x49} 
\end{eqnarray} 

In Figure~\ref{fig2} we compare the behaviours of the state of the system as 
represented by the Wigner-Stratonovic function under the Lindblad dynamics 
and the POVM iteration for an initial coherent state of $J=1$ and $J=5$ spin 
systems. These correspond to the initial states used in Figure~\ref{fig1}. Note 
the different colour scales used for the different plots. We remark on the 
observation that the spin $J=5$ coherent state ``decoheres'' slower than the 
spin $J=1$ state in both POVM and Lindblad evolution, if we interpret 
decoherence as merely the decay of the matrix elements. Also note that while 
both dynamics asymptotically lead the state to the uniform distribution on the 
sphere, there are marked differences in the fluctuations between POVM and 
Lindblad evolution, where the Lindblad evolution washes out the features of 
the Wigner function faster than the POVM iterations. These differences are, 
as expected, more pronounced for the smaller value of $J$. 

To contrast the impact of decoherence on an initial coherent state, in 
Figure~\ref{fig3} we take a highly quantum initial state 
\begin{eqnarray}
|\psi\rangle= \frac{1}{\sqrt{2}}\left(|m = J\rangle+|m = -J\rangle\right) 
\end{eqnarray} 
and compare the effect of the two decoherence models on this state. 
Here, the left column shows the Wigner 
function of the initial state (they are identical). In the middle and the right 
columns, the top row shows the resulting Wigner function under the POVM 
iterations (for $n=1,5$), while the bottom row shows the 
solution to the Lindblad equation with $\gamma=1/J$ at $t=1,5$. Once the 
interference fringes are washed out, the application of POVM measurements 
appears to keep the structure of the Wigner function. Note, however, that the 
colour scales are changing, indicating an approach to a uniform distribution 
while keeping the structure. Under the Lindbladian dynamics (bottom row), 
on the other hand, the approach to uniformity is more pronounced.

\section{Decoherence timescale and positivity of the Wigner function} 
\label{sec:6} 

In the context of phase-space representations, the positivity of the 
quasidistribution function is often considered in the literature. One reason for 
the interest on positivity is that if $F^\sigma$ for $\sigma=+1$, the 
Glauber-Sudarshan $P$ function, is nonnegative, then the state ${\hat\rho}$ 
is often viewed as being classical \cite{Braun}. On the other hand,
we recall that the quasidistribution function for $\sigma=-1$ is the Husimi 
density, which is strictly nonnegative. It follows from (\ref{eq:x49}) 
that for any value of $\sigma$, 
the quasidistribution function is transformed into a positive density function 
after applying at most $\lceil (\sigma + 1)/2 \rceil$ iterations of phase-space 
measurements. (See \cite{BM} for an analogous analysis on tomographic 
decoherence models.)
This provides a sharp upper bound on the number of 
measurements required to guarantee transition into classicality, if classicality 
were to be interpreted as the positivity of a given quasidistribution function. 

An analogous question on the positivity of $F^\sigma$ can be addressed 
under the Lindblad dynamics (\ref{eq:1}), as discussed also in \cite{DM}. Here 
we remark that for $J=1/2$, owing to the equivalence of the two decoherence 
models, we are able to give an exact time at which a given quasidistribution 
function becomes positive. This follows on account of the fact that the solution 
to the Lindblad dynamics (\ref{eq:1}) agrees with the resulting state after $n$ 
iterations of the POVM measurements at times 
\begin{eqnarray}
t_n = \frac{\log 3}{\gamma} \, n \, . 
\end{eqnarray} 
It follows that for $J=1/2$ and any $\sigma$, the quasidistribution 
function becomes positive under the Lindblad dynamics at time 
\begin{eqnarray}
t_\sigma^* = \frac{\log 3}{\gamma} \,\lceil (\sigma + 1)/2 \rceil \, ,
\end{eqnarray} 
which is an exact result. For $J>1/2$ we do not have an exact result, but 
it seems reasonable to conjecture that positivity is ensured provided that 
the damped $\sigma$ kernel in (\ref{eq:x44}), due to decoherence, becomes 
no sharper than the Husimi ($\sigma=-1$) kernel. That is, if 
\begin{eqnarray}
\re^{-\frac{1}{2}\gamma L(L+1)} \left(\frac{{2J \choose L}}
{{2J+L+1 \choose L}}\right)^{-\frac{\sigma}{2}} \leq 
\left(\frac{{2J \choose L}}{{2J+L+1 \choose L}}\right)^{\frac{1}{2}}  
\end{eqnarray} 
for all $L$, then we restore positivity. 
Now the largest value the binomial ratio can take is for $L=2J$, 
so solving this for $t$ we find that 
\begin{eqnarray}
t_\sigma^* = \frac{\sigma+1}{2\gamma J(2J+1)} \,
\log {4J+1 \choose 2J}  \, .
\end{eqnarray} 
In particular, for $J\gg1$ this gives 
\begin{eqnarray}
t_\sigma^* \approx \frac{\sigma+1}{4\gamma J^2} \left( 4J \log2 - \half \log(2\pi J) 
\right)  \, .
\end{eqnarray} 
This result suggests that if $\gamma$ is held fixed for different $J$, then for 
larger spins the positivity is attained faster, according to $t_\sigma^*\sim 1/J$. 
On the other hand, should $\gamma$ scale like $\gamma=1/J$, then we 
have $t_\sigma^* \approx(\sigma+1)\log2$.

\section{Summary and Discussion} 
\label{sec:7} 

In summary, we have presented two models for phase-space decoherence 
of spin systems, one based on a homogeneous spin dephasing induced by the 
Lindblad dynamics, and one based on the phase-space POVM measurements. 
We have shown that for $J>1/2$ the associated decoherence rates are not in 
agreement, although the qualitative behaviours of the system implied by the 
two decoherence models are very similar and the discrepancy rapidly 
disappears in increasing $J$. 
We found, perhaps not surprisingly, that a coherent state with an increasing 
$J$ becomes more robust against phase-space decoherence. 
We have also compared the two decoherence models in terms of the 
phase-space quasidistribution functions, and provided estimates for the first 
passage times for their positivity.  

In this connection it is worth drawing attention to the fact that 
there are two different timescales that emerge, namely, the timescale for the 
decay of the elements of the density matrix, and the timescale for recovering 
the positivity of the quasiprobability distributions. Traditionally, decoherence is 
viewed as the decay effect of the elements of the density matrix, and here we 
find that the decay rates are smaller for larger spins. On the other hand, 
decoherence is 
often viewed as the signature for the emergence of classicality, but so is the 
positivity and the disappearance of interference fringes in the quasiprobability 
distributions \cite{Mann2021}. 
However, for the latter, the timescale is smaller for larger spins, 
which is the opposite of the decay rate analysis. Hence depending on which 
criterion (decay of the density matrix or the positivity of the quasiprobability 
distribution) one chooses, the answer to the question whether systems with 
larger spins classicalise faster differs.

\vspace{0.2cm} 
\section*{Acknowledgements}
RM acknowledges support through an Imperial College President's PhD Scholarship. 

\appendix 
\section{Microscopic derivation of the Lindblad equation (\ref{eq:1})}\label{app_LB}
If three independent components of the spin or angular momentum of a particle 
are randomly monitored by the environment, then we may model the interaction 
between the system, whose state at time $t$ is given by ${\hat\rho}_t$, with its 
environment over the time period $[t,t+\rd t]$ through a unitary operator by writing 
\begin{eqnarray}
{\hat\rho}_t + \rd {\hat{\boldsymbol\rho}}_t &=& 
\re^{-{\rm i}\sqrt{\gamma}({\hat J}_x w^x_t + 
{\hat J}_y w^y_t + {\hat J}_z w^z_t){\rm d}t} 
\nonumber \\ && \quad \times 
\,{\hat\rho}_t\, \re^{{\rm i}\sqrt{\gamma} 
({\hat J}_x w^x_t + {\hat J}_y w^y_t + {\hat J}_z w^z_t){\rm d}t} 
\label{eq:SM1} 
\end{eqnarray}
for an infinitesimal modification of the state over a short time period $\rd t$. Here, 
we represent the three independent components of the angular momentum by the 
operators $\{{\hat J}_x,{\hat J}_y,{\hat J}_z\}$, and we use a bold font 
$\rd{\hat{\boldsymbol\rho}}_t$ for the state transformation to highlight the fact that 
it is a random perturbation that will have to be averaged to recover the density 
matrix. The interaction with the 
environment, however, is highly oscillatory and is thus random, and we model this 
by use of three statistically-independent Gaussian white noise terms $\{w_t^x, w_t^y, 
w_t^z\}$. We let $\sqrt{\gamma}$ model the coupling strength. In essence, our 
microscopic model takes the form of the Peres model for the system-environment 
interaction for the monitoring of the system \cite{Peres}. For $\rd t\ll1$ we expand 
(\ref{eq:SM1}) to obtain 
\begin{eqnarray}
{\hat\rho}_t + \rd {\hat{\boldsymbol\rho}}_t &=& \left( 1 - \ri \sqrt{\gamma} (\hat{\boldsymbol J}\!
 \cdot \! {\boldsymbol w}_t)\rd t - \half \gamma (\hat{\boldsymbol J}\! \cdot \! 
{\boldsymbol w}_t)^2 (\rd t)^2\right)  {\hat\rho}_t  
\nonumber \\ && \times 
\left( 1 + \ri \sqrt{\gamma} 
(\hat{\boldsymbol J}\! \cdot \! {\boldsymbol w}_t)\rd t - \half \gamma 
(\hat{\boldsymbol J}\! \cdot \! {\boldsymbol w}_t)^2 (\rd t)^2 \right) , 
\nonumber \\ 
\label{eq:SM2} 
\end{eqnarray}
where we have written $\hat{\boldsymbol J}\! \cdot \! {\boldsymbol w}_t$ for 
${\hat J}_x w^x_t + {\hat J}_y w^y_t + {\hat J}_z w^z_t$. Expanding the 
right side of (\ref{eq:SM2}), making use of the property $(w_t \, \rd t)^2=\rd t$ 
of the Gaussian white noise \cite{Hida}, and taking the 
average of the resulting expression for $\rd {\hat{\boldsymbol\rho}}_t$, 
noting that independent Gaussian white noise terms have vanishing expectations, 
and writing $\rd{\hat\rho}_t={\mathds E}[\rd {\hat{\boldsymbol\rho}}_t]$ for the 
expectation, we deduce the Lindblad equation (\ref{eq:1}).

\onecolumngrid
\newpage

\end{document}